\documentclass[10pt]{IEEEtran} 

\usepackage{graphicx,color}
\usepackage{dcolumn}
\usepackage{bm}
\usepackage{mathbbol}
\usepackage{amsmath,amssymb}  
\usepackage[latin9]{inputenc} 

\usepackage[sans]{dsfont}
\usepackage[mathscr]{euscript}

\usepackage{amssymb}          
\usepackage{amsmath}          

\newtheorem{prop}{Proposition}
\newtheorem{defin}{Definition}
\newtheorem{thm}{Theorem}
\newtheorem{cor}{Corollary}
\newtheorem{lemma}{Lemma}
\newtheorem{prob}{Problem}

\newcommand{\proof}{\noindent {\em Proof. }}

\newcommand{\ket}[1]{|#1\rangle}
\newcommand{\bra}[1]{\langle #1|}

\newcommand{\supp}{\rm{supp}}

\newcommand{\beq}{\begin{equation}}
\newcommand{\eeq}{\end{equation}}
\newcommand{\beqa}{\begin{eqnarray}}
\newcommand{\eeqa}{\end{eqnarray}}
\newcommand{\beqan}{\begin{eqnarray*}}
\newcommand{\eeqan}{\end{eqnarray*}}

\newcommand{\trace}{{\rm Tr}}

\newcommand{\tr}[2]{\text{Tr}_{#1}\left(#2\right)}

\renewcommand{\ket}[1]{| #1 \rangle}
\renewcommand{\bra}[1]{\langle #1 |}

\newcommand{\ketbra}[1]{\ket{#1}\bra{#1}}

\newcommand{\identity}{\mathbb{I}}
\newcommand{\hilbert}{\mathcal{H}}

\newcommand{\neigh}{\mathcal{N}}

\newcommand{\Hi}{\mathcal{H}}

\renewcommand{\P}{\mathbb{P}}
\newcommand{\C}{\mathbb{C}}

\newcommand{\DD}{{\mathfrak D}(\Hi)}

\newcommand{\Li}{\mathcal{L}}
\newcommand{\Si}{\mathcal{S}}
\newcommand{\Ei}{\mathcal{E}}
\newcommand{\Fi}{\mathcal{F}}
\newcommand{\Ai}{\mathcal{A}}

\newcommand{\Ni}{\mathcal{N}}
\newcommand{\alg}{{\rm alg}}
\newcommand{\fix}{{\rm fix}}

\renewcommand{\supp}{{\rm supp}}

\renewcommand{\ker}{{\rm ker}}
\renewcommand{\span}{{\rm span}}

\newcommand{\Tr}{{\rm Tr}}

\newcommand{\red}{\color{black}}

\newcommand{\qed}{\hfill $\Box$ \vskip 2ex}

\begin{document}

\title{Alternating Projections Methods for\\ Discrete-time Stabilization of Quantum States}

\author{Francesco Ticozzi\thanks{F. Ticozzi is with the Dipartimento di Ingegneria dell'Informazione,
Universit\`a di Padova, via Gradenigo 6/B, 35131 Padova, Italy, and with the Department of Physics and Astronomy, 
Dartmouth College, 6127 Wilder Laboratory, Hanover, NH 03755, USA (email: {ticozzi@dei.unipd.it}).},
Luca Zuccato\thanks{L. Zuccato is with the Dipartimento di Ingegneria dell'Informazione,
Universit\`a di Padova, via Gradenigo 6/B, 35131 Padova, Italy (email: {zuccato@studenti.unipd.it}).},
Peter D. Johnson\thanks{P. D. Johnson did this work while at the Department of Physics and Astronomy,
Dartmouth College, 6127 Wilder Laboratory, Hanover, NH 03755, USA. Current address: 
Department of Chemistry and Chemical Biology
Harvard University, 12 Oxford Street, Cambridge, MA 02138, USA Ê
(email: {peter.d.johnson22@gmail.com}).},
Lorenza Viola\thanks{L. Viola is with the Department of Physics and Astronomy,
Dartmouth College, 6127 Wilder Laboratory, Hanover, NH 03755, USA (email: {lorenza.viola@dartmouth.edu}).}}

\maketitle

{
\abstract            
We study sequences (both cyclic and randomized) of idempotent completely-positive 
trace-preserving quantum maps, and show how they asymptotically converge to the intersection of their fixed 
point sets via alternating projection methods.
We characterize the robustness features of the protocol against randomization and provide basic bounds on its 
convergence speed. The general results are then specialized to stabilizing entangled states in {\red finite-dimensional} 
multipartite quantum systems subject to a resource constraint, a problem of key interest for quantum information applications. 
We conclude by suggesting further developments, including techniques to enlarge the set of 
stabilizable states and ensure efficient, finite-time preparation. 
}

\section{Introduction}

Driving a quantum system to a desired state is 
a prerequisite for quantum control applications ranging from 
quantum chemistry to quantum computation \cite{altafini-tutorial}. 
An important distinction arises depending on whether the target state is to be obtained starting 
from a {\em known} initial state -- in which case such knowledge may be leveraged to design
a control law that effects a ``one-to-one'' state transfer; or, the initial state is 
{\em arbitrary} (possibly unknown) -- in which case the dynamics must allow for ``all-to-one'' transitions. 
We shall refer to the latter as a {\em state preparation} task.
The feasibility of these tasks, as well as the robustness and efficiency of the control protocol 
itself, are heavily influenced by the control resources that are permitted. Unitary control can 
only allow for transfer between states (represented by density operators),  that are 
iso-spectral \cite{Schirmer2002}.  
For both more general quantum state transfers, and for all state preparation tasks,
access to {\em non-unitary} control (in the form of either coupling to an external 
reservoir or employing measurement and feedback) becomes imperative. 

If a dissipative mechanism that ``cools'' the system to a known pure state is available, 
the combination of this all-to-one initialization step with state-controllable one-to-one unitary dynamics 
\cite{dalessandro-book} is the simplest approach for achieving pure-state preparation, 
with methods from optimal control theory being typically 
employed for synthesizing the desired unitary dynamics \cite{altafini-tutorial,Giovanna2014}.
In the same spirit, in the circuit model of quantum computation \cite{nielsen-chuang}, 
preparation of arbitrary pure states 
is attained by initializing the quantum register in a known factorized pure state, and then 
implementing a sequence of unitary transformations (``quantum gates'') drawn from a universal set. 
Sampling from a mixed target state can be obtained by allowing for randomization of the 
applied quantum gates in conjunction with Metropolis-type algorithms \cite{quantum-metropolis}. 
Additional possibilities for state preparation arise if the target system is allowed to couple to 
a quantum auxiliary system, so that the pair can be jointly initialized and controlled, and the 
ancilla reset or traced over  \cite{viola-engineering,ticozzi-cooling}.
For example, in \cite{wolf-sequential} it is shown how sequential unitary coupling to an ancilla may be used to 
design a sequence of non-unitary transformations (``quantum channels'') on a target multi-qubit system
that dissipatively prepare it in a matrix product state. 

{In the above-mentioned state-preparation methods, the  dissipative mechanism is {\em fixed} (other than being turned on 
and off as needed), and the control design happens at the unitary level, directly on the target space or an enlarged one. 
A more powerful setting is to 
allow {\em dissipative control design} from the outset \cite{poyatos,viola-engineering,kraus-controllability}. This opens 
up the possibility to synthesize all-to-one open-system dynamics that not only prepare the target state of interest 
but, additionally, leave it {\em invariant} throughout -- that is, achieves {\em stabilization}, which is the task we focus on in 
this work.  
Quantum state stabilization has been investigated from different perspectives, including feedback 
design with classical \cite{belavkin-feedback,wiseman-feedback,doherty-linear,vanhandel-feedback,mirrahimi-stabilization, rouchon-nature, ticozzi-QDS, ticozzi-markovian} and quantum \cite{wiseman-milburn,james-commmath,wang-coherent,baggio-CDC} controllers, 
as well as open-loop reservoir engineering techniques with both time-independent dynamics and switching control 
\cite{davidovich,polzik-dissipative,ticozzi-NV,ticozzi-ql,ticozzi-ql1,mirrahimi-nature, ticozzi-generic,kraus-dissipative,scaramuzza-switching}.
Most of this research effort, however, has focused on continuous-time models, 
with fewer studies addressing {\em discrete-time} quantum dynamics. With ``digital'' open-system quantum simulators 
being now experimentally accessible \cite{barreiro,blatt-maps}, {investigating 
quantum stabilization problems} in discrete time becomes both natural and important. 
Thanks to the invariance requirement, ``dissipative quantum circuits'' bring distinctive advantages
toward preparing pure or mixed target states {\em on-demand}, notably:

\begin{itemize}
\item While in a unitary quantum circuit or a sequential generation scheme, the desired state is only 
available at the completion of the full protocol, invariance of the target ensures that 
{repeating a stabilizing protocol or even portions of it, 
will further maintain the system in the target state (if so desired), without disruption.}

\item The order of the applied control operations need no longer be crucial, 
allowing for the target state to still be reached probabilistically (in a suitable sense), while 
incorporating robustness against the implementation order.

\item If at a certain instant a wrong map is implemented, or some transient noise perturbs the dynamics, 
these unwanted effects can be re-absorbed without requiring active intervention or the whole 
preparation protocol having to be re-implemented correctly. 

\end{itemize}

Discrete-time quantum Markov dynamics are described by sequences of quantum channels, namely, 
completely-positive, trace-preserving (CPTP) maps \cite{kraus}. This
give rise to a rich stability theory that can be seen as the non-commutative generalization of 
the asymptotic analysis of classical Markov chains, 
and that thus far has being studied in depth only in the time-homogeneous case \cite{wolf-notes,cirillo-decompositions}, 
including elementary feedback stabilizability and reachability problems
\cite{Bolognani2009,albertini-feedback}. 

In this work, we show that {\em time-dependent} sequences of CPTP maps can be used to make their common fixed states the 
minimal asymptotically stable sets, which are reached by iterating cyclically a finite subsequence. 
The methods we introduce 
employ a finite number of {idempotent} CPTP maps, which we call {\em CPTP projections}, and can be considered a quantum version of {\em alternated projections methods}. 
The latter, stemming from seminal results by von Neumann \cite{neumann1950functional} and extended by Halperin \cite{halperin1962product} and others \cite{bauschke,escalante}, are a family of (classical) algorithms that, loosely speaking, aim to select an element in the intersection of a number of sets that minimizes a natural (quadratic) distance with respect to the input. 
The numerous applications 
of such classical algorithms include estimation \cite{dykstra1983algorithm} and control \cite{grigoriadis1994fixed} and, recently, 
specific tasks in quantum information, such as quantum channel construction \cite{quantumprojection2015}. 
In the context of quantum stabilization, we show that instead of working with the standard (Hilbert-Schmidt) inner product,
it is natural to resort to a different inner product, a {\em weighted inner product} for which the CPTP projections become orthogonal, 
and the original results apply. When, depending on the structure of the fixed-point set, this strategy is not viable, we establish convergence by a different proof that does not directly build on existing alternating projection theorems. For all the proposed 
sequences, the order of implementation is not crucial, and convergence in probability is guaranteed even when the sequence is randomized, under very mild hypotheses on the distribution. 
 
Section \ref{sec:preliminary} introduces the models of interest, 
and recalls some key results regarding stability and fixed points of CPTP maps.
Our general results on quantum alternating projections are presented in Section \ref{sec:alternatingprojections}, after proving that 
CPTP projections can be seen as orthogonal projections with respect to a weighted inner product. Basic bounds on the 
speed of convergence and robustness of the algorithms are discussed in Sections \ref{sec:speed} and \ref{sec:robustness}, respectively.
In Section \ref{sec:ql} we specialize these results to distributed stabilization of entangled states on multipartite quantum systems, where the {robustness} properties imply that the target can be reached by {\em unsupervised randomized applications} of dissipative quantum maps.

\section{Preliminary material}\label{sec:preliminary}

\subsection{Models and stability notions}

We consider a finite-dimensional quantum system, associated to a Hilbert space $\Hi\approx\C^d.$ Let ${\cal B}(\Hi)$ denote the 
space of linear bounded operators on $\Hi,$ with $\dag$ being the adjoint operation.
The state of the system at each time $t \geq 0$ is a {\em density matrix} in $\DD,$ namely a positive-semidefinite, 
trace one matrix. Let $\rho_0$ be the initial state. 
We consider {\em time inhomogenous} Markov dynamics, namely, sequences of CPTP maps $\{\Ei_t\},$ defining the 
state evolution for  through the following dynamical equation:
\beq\rho_{t+1}=\Ei_t(\rho_t), \quad t\geq0.
\eeq
Recall that a linear map $\Ei$ is CPTP if and only if it admits an operator-sum representation (OSR) \cite{kraus}:
\[\Ei(\rho)=\sum_k M_k\rho M_k^\dag,\]
where the (Hellwig-Kraus) operators $\{M_k\}\subset {\cal B}(\Hi)$ satisfy $\sum_k M_k^\dag M_k=I.$ 
We shall assume that for all $t>0$ the map $\Ei_t=\Ei_{j(t)}$ is chosen from a set of ``available'' maps, to be designed 
within the available control capabilities. In particular, in Section \ref{sec:ql} we will focus on locality-constrained dynamics.
For any {$t \geq s\geq 0,$} we shall denote by
\[\Ei_{t,s} \equiv \Ei_{t-1}\circ\Ei_{t-2}\circ \ldots\circ \Ei_s, \quad (\Ei_{t,t}= {\cal I}), \]
the evolution map, or ``propagator'', from $s$ to $t.$

A set $\Si$ is {\em invariant} for the dynamics if $\Ei_{t,s}(\tau)\in\Si$ for all $\tau\in\Si$. 
Define the distance of an operator $\rho$ from a set $\Si$ as 
\[d(\rho,\Si) \equiv \inf_{\tau\in\Si}\|\rho-\tau\|_1,\]
with $\|\cdot\|_1$ being the trace norm.
The following definitions are straightforward adaptations of the standard ones \cite{khalil}: 


\begin{defin}
(i) An invariant set $\Si$ is (uniformly) {\em simply stable} if for any $\varepsilon>0$ there exists $\delta > 0$ such that $d(\tau,\Si)<\delta$ ensures $d(\Ei_{t,s}(\tau),\Si)<\varepsilon$ for all $t\geq s\geq 0.$  
(ii) An invariant set $\Si$ is {\em globally asymptotically stable} (GAS) if it is simply stable and
\beqa 
&&\lim_{t\rightarrow\infty}d(\Ei_{t,s}(\rho),\Si)=0,\quad \forall \rho,\; s\geq 0.\label{patt}\eeqa 
\end{defin}

Notice that, since we are dealing with finite-dimensional systems, convergence in any matrix norm is equivalent. 
Furthermore, since CPTP maps are trace-norm contractions \cite{nielsen-chuang}, we have that simple stability is always 
guaranteed (and actually the distance is monotonically non-increasing):

\begin{prop} 
If a set $\Si$ is invariant for the dynamics $\{\Ei_{t,s}\}_{t,s\geq 0}$, then it is simply stable.
\end{prop}
\proof
We have, for all ${t,s\geq 0}$:
\beqan d(\Ei_{t,s}(\rho),\Si)&\leq& d(\Ei_{t,s}(\rho),\Ei_{t,s}(\tau_{t,s}^*))\\
&\leq& d(\rho,\tau_{t,s}^*) \\
&=&d(\rho,\Si).
\eeqan
The first inequality is true, by definition, for all $ \tau_{t,s} \in\Si,$ and also on the closure $\bar \Si$, thanks to continuity of $\Ei_{t,s}$; 
the second holds due to contractivity of $\Ei,$ and the last equality follows by letting  
$\tau_{t,s}^* \equiv {\rm arg}\min_{\tau\in\bar\Si}\|\rho-\tau\|_1,$ where we can take the min since $\bar\Si$ is closed and compact.
\qed


\subsection{Fixed points of CPTP maps}

We collect in this section some relevant results on the structure of fixed-point sets 
for a CPTP map $\Ei$, denoted by $\fix(\Ei).$ 
More details can be found in \cite{johnson-ffqls,petz-book,wolf-notes,viola-IPSlong}.

Let $\alg(\Ei)$ denote the $\dag$-closed algebra generated by the operators in the 
OSR of $\Ei,$ and $\Ai'$ denote the commutant of $\Ai$, respectively. 
For {\em unital} CP maps, $\fix(\Ei)$ is a $\dag$-closed algebra, $\fix(\Ei)=\alg(\Ei)'=\fix(\Ei^\dag)$ 
\cite{wolf-notes,viola-IPSlong}. This implies that it admits a (Wedderburn) block decomposition \cite{davidson}:
\beq
\label{wedd}
\fix(\Ei) = \bigoplus_\ell {\cal B}(\Hi_{S,\ell}) \otimes I_{F,\ell}, 
\eeq
with respect to a Hilbert space decomposition:
\[\Hi=\bigoplus_\ell \Hi_{S,\ell} \otimes \Hi_{F,\ell}.\]

\noindent 
For general (not necessarily unital) CPTP maps the following holds 
\cite{viola-IPSlong,wolf-notes,johnson-ffqls}: 

\begin{thm}[Fixed-point sets, generic case]
\label{thm:dualkernelkraus}
\label{thm:generalfixedpointkraus}
Given a CPTP map $\Ei$ which admits a full-rank fixed point $\rho$, we have
\begin{equation}
\fix(\Ei)=\rho^\frac{1}{2}\,\fix(\Ei^\dag)\,\rho^\frac{1}{2}. 
\end{equation}
Moreover, with respect to the decomposition of $\fix(\Ei^\dag)=
\bigoplus_\ell {\cal B}(\Hi_{S,\ell}) \otimes {I}_{F,\ell}$, 
{the fixed state has the structure:}
\beq\label{rhostr}\rho = \bigoplus_\ell p_\ell\rho_{S,\ell}\otimes \tau_{F,\ell},
\eeq
where $\rho_{S,\ell}$ and $\tau_{F,\ell}$ are {\em full-rank} density operators of appropriate dimension, 
and $p_\ell$ a set of convex weights.
\end{thm}

This means that, given a CPTP map
admitting a full-rank invariant state $\rho$, the fixed-point sets $\fix(\Ei)$ is a 
{\em $\rho$-distorted algebra}, namely, an associative algebra with respect to a modified product 
(i.e. $X \times_\rho Y= X\rho^{-1} Y$), 
with structure
\beq
\label{CPTPfixedpoints}
{\cal A}_\rho=\bigoplus_\ell {\cal B}(\Hi_{S,\ell}) \otimes \tau_{F,\ell}, \eeq
where $\tau_{F,\ell}$ are a set of density operators of appropriate dimension 
(the same for every element in $\fix(\Ei)$). 
In addition, since $\rho$ has the same block structure \eqref{rhostr}, 
$\fix(\Ei)$ is clearly invariant with respect to the action of the linear map 
${\cal M}_{\rho,\lambda}(X) \equiv  \rho^{\lambda} X \rho^{-\lambda}$ for any $\lambda \in \mathbb{C}$, and 
in particular for the {\em modular group} $\{{\cal M}_{\rho,i\phi}\}$ \cite{petz-book}. The same holds for the fixed points of the dual dynamics. 
In \cite{johnson-ffqls}, the following result has been proved using Takesaki's theorem, showing that commutativity with a modular-type operator is actually sufficient to ensure that a distorted algebra is a valid fixed-point set.

\begin{thm}
\label{thm:modinvariance}
{\em(Existence of $\rho$-preserving dynamics)}
Let $\rho$ be a full-rank density operator. A distorted algebra ${\cal A}_\rho,$ such that $\rho\in{\cal A}_\rho,$ admits a CPTP map $\Ei$ such that $\fix(\Ei)={\cal A}_\rho$ if and only if it is invariant for ${\cal M}_{\rho,\frac{1}{2}}.$
\end{thm}

To our present aim, it is worth remarking that in the proof of the above result, a CPTP idempotent map is derived as the dual of a {\em conditional expectation} map, namely, the orthogonal projection onto the (standard) algebra $\fix(\Ei^\dag).$

If the CPTP map $\Ei$ does not admit a full rank invariant state, then it is possible to characterize the fixed-point set by first reducing to the support of the invariant states. This leads to the following structure theorem 
\cite{viola-IPSlong,wolf-notes,johnson-ffqls}:

\begin{thm}[Fixed-point sets, general case]
\label{thm:dualkernel-general}  
Given a CPTP map $\Ei,$ 
and a maximal-rank fixed point $\rho$ with $\tilde\Hi\equiv \supp(\rho)$, let 
$\tilde\Ei$ denote the reduction of $\Ei$ to $\mathcal{B}(\tilde\Hi)$.  We then have
\begin{equation}
\fix(\Ei)=\rho^\frac{1}{2}\,(\ker(\tilde\Ei^\dag)\oplus {\mathbb O}) \,\rho^\frac{1}{2}, 
\end{equation}
where ${\mathbb O}$ is the zero operator on the complement of $\tilde \Hi.$
\end{thm}

\section{Alternating projection methods}
\label{sec:alternatingprojections}

\subsection{von Neumann-Halperin Theorem}

Many of the ideas we use in this paper are inspired by a classical result originally due to von Neumann 
\cite{neumann1950functional}, and later extended by Halperin to multiple projectors:

\begin{thm}
[\hspace*{-0.5mm}von Neumann-Halperin alternating projections] 
\label{thm:halp} 
If $\mathcal{M}_1$,\ldots,$\mathcal{M}_r$ are closed subspaces in a Hilbert space $\hilbert$, and 
$P_{\mathcal{M}_j}$ are the corresponding orthogonal projections, then 
 $$\lim_{n\rightarrow\infty}(P_{\mathcal{M}_1}...P_{\mathcal{M}_r})^nx=Px, \quad \forall x\in \hilbert, $$
where $P$ is the orthogonal projection onto ${\bigcap_{i=1}^r\mathcal{M}_i}$.
\end{thm}

A proof for this theorem can be found in Halperin's  original work \cite{halperin1962product}. Since then, the result 
has been refined in many ways, has inspired similar convergence results that use {information projections 
\cite{csiszar-divergence}} and, in full generality, projections in the sense of Bregman divergences \cite{bregman,bauschke}.
The applications of the results are manifold, especially in algorithms: while it is beyond the scope of this work to attempt a review, a good collection is presented in \cite{escalante}.
Some bounds on the convergence rate for the alternating projection methods can be derived by looking at the 
{\em angles between the subspaces} we are projecting on.  We recall their definition and basic properties in Appendix \ref{app:cosine}, see again \cite{escalante} for more details.

\subsection{CPTP projections and orthogonality}

We call an idempotent CPTP map, namely, one that satisfies $\mathcal{E}^2=\mathcal{E}$, a {\em CPTP projection.} 
As any linear idempotent map, $\Ei$ has only $0,1$ eigenvalues and maps any operator $X$ onto the set of its fixed 
points, $\fix(\mathcal{E})$. Recall that 
\begin{equation}
\label{eq:dec_fix}
\fix(\mathcal{E})=\bigoplus_\ell  [ \mathcal{B}(\hilbert_{S,\ell})\otimes \tau_{F,\ell} ] \oplus \mathbb{O}_R,
\end{equation} 
for some Hilbert-space decomposition:
\begin{equation}
\label{eq:decomposition}
\hilbert= \bigoplus_\ell (\hilbert_{S,\ell}\otimes \hilbert_{F,\ell})\oplus\Hi_R,
\end{equation}
where the last zero-block is not present if there exists a $\rho>0$ in $\fix(\Ei).$
{We next give the structure of the CPTP projection 
associated to $\fix(\mathcal{E})$:} 
The result is essentially known (see e.g. \cite{wolf-notes,petz-book}) but we provide a short proof for completeness):
{\begin{prop} 
\label{prop:proj} 
Given a CPTP map $\mathcal{E}$ with $\rho$ a fixed point of maximal rank, 
a CPTP projection onto $\Ai_\rho=\fix (\Ei)$ exists and is given by
\beq
\label{einfty} 
\Ei_{\Ai_\rho}(X)=\lim_{T\to+\infty}\frac{1}{T}\sum_{i=0}^{T-1}\Ei^i(X).
\eeq
If the fixed point $\rho$ is full rank, then the CPTP projection onto $\Ai_\rho = \oplus_\ell 
{\cal B}(\Hi_{S,\ell}) \otimes \tau_{F,\ell} $ is equivalently given by
\begin{equation}
\label{eq:proj_2}
{\mathcal{E}_{\Ai_\rho}}(X)=\bigoplus_\ell
\trace_{F,\ell}(\Pi_{SF,\ell}X\, \Pi_{SF,\ell})\otimes {{\tau_{F,\ell}}},  
\end{equation}
where $\Pi_{SF,\ell}$ is the orthogonal projection from $\Hi$ onto the subspace 
$\hilbert_{S,\ell} \otimes \hilbert_{F,\ell}$. 
\end{prop}}

\proof 
Recalling that $\|\Ei\|_\infty=1$, it is straightforward to prove that the limit in Eq. \eqref{einfty} 
exists and is a CPTP map. Furthermore, clearly $\fix(\Ei)\subseteq\fix(\Ei_{{\cal A}_\rho}),$ and 
\[\Ei_{{\cal A}_\rho}\Ei=\Ei\Ei_{{\cal A}_\rho}=\Ei_{{\cal A}_\rho}.\]
It follows that $\Ei_{{\cal A}_\rho}^2=\Ei_{{\cal A}_\rho}.$  On the other hand, it is immediate to verify that the right hand side of Eq. \eqref{eq:proj_2} is CPTP, has image equal to its fixed points ${\mathcal{A}}_\rho= \fix({\mathcal{E}}),$ and is idempotent. Hence, it coincides with $\Ei_{\Ai_\rho}.$ 
\qed

{For a full-rank fixed-point set, CPTP projections are {\em not} orthogonal projections onto $\fix(\Ei)$, 
at least with respect to the Hilbert-Schmidt inner product.
The proof is given in Appendix \ref{app1}. We are nonetheless going to show that $\Ei_\Ai$ {\em is} an orthogonal projection 
with respect to a different  inner product. This proves that the map in Eq. \eqref{eq:proj_2} is the unique CPTP projection onto  $\Ai_\rho$.
If the fixed-point set does not contains a full-rank state,  Eq. \eqref{einfty} still defines a valid CPTP projection onto 
$\fix(\Ei)$; however, this need not be unique.} 
We will exploit this fact in the proof of Theorem \ref{thm:alternating_pure}, 
where we choose a particular one. 

{\defin 
\label{def:mod_in_prod} 
Let $\xi$ be a positive-definite operator. (i) We define the {\em $\xi$-inner product} as
\beq 
\label{eq:md_in_prod}
\langle X,Y\rangle_\xi \equiv \trace(X\xi Y);\eeq
(ii) We define the {\em symmetric $\xi$-inner product} as
\beq 
\label{eq:md_sym} 
\langle X,Y\rangle_{\xi,s} \equiv \trace(X\xi^\frac{1}{2}Y\xi^\frac{1}{2}).
\eeq}
\noindent 
It is straightforward to verify that both \eqref{eq:md_in_prod} and \eqref{eq:md_sym} are valid inner products.

We next show that $\mathcal{E}_\Ai$ is an orthogonal projection with respect to \eqref{eq:md_in_prod} and \eqref{eq:md_sym}, when $\xi=\rho^{-1}$ for a full rank fixed point $\rho.$ We will need a preliminary lemma. 
With $W \equiv \bigoplus W_i$ we will denote an operator that acts as $W_i$ on $\hilbert_i$, for a direct-sum 
decomposition of $\hilbert=\bigoplus_i \hilbert_i$. 

{\lemma \label{lemma:blocks} Consider  $Y,W\in\mathcal{B}(\Hi),$ where $W$ admits an orthogonal block-diagonal representation $W=\bigoplus_\ell W_\ell$. Then $\trace(WY)=\sum_\ell\trace( W_\ell Y_\ell)$, where $Y_\ell=\Pi_\ell Y\Pi_\ell$. }

{\proof Let $\Pi_\ell$ be the projector onto $\hilbert_\ell$. Remembering that $\sum_\ell\Pi_\ell=I$ and $\Pi_\ell=\Pi_\ell^2$, it  follows that 
\begin{eqnarray*}
 \trace(X) = \sum_\ell \trace(\Pi_\ell X)
        =\sum_\ell \trace(\Pi_\ell X\Pi_\ell).
\end{eqnarray*} 
Therefore, we obtain:
\begin{eqnarray*}
\trace(WY)&=&\trace(\sum_\ell\Pi_\ell \bigoplus_j W_j Y)=\sum_\ell\trace( \Pi_\ell W_\ell  Y)\\
          &=&\sum_\ell\trace( \Pi_\ell W_\ell\Pi_\ell  Y)=\sum_\ell\trace(  W_\ell \Pi_\ell Y\Pi_\ell)\\
          &=&\sum_\ell\trace( W_\ell Y_\ell). \hspace{4.2cm} \Box
\end{eqnarray*}
}
{\begin{prop} 
\label{prop:orth_proj_mod_in} Let $\xi=\rho^{-1}$, where $\rho$ is a full-rank fixed state in $\Ai_\rho$, which is invariant for ${\cal M}_{\rho,\frac{1}{2}}.$ Then $\Ei_{\Ai_\rho}$ is an orthogonal projection with respect to the inner products in  \eqref{eq:md_in_prod} and \eqref{eq:md_sym}.
\end{prop}}

{\proof We already know that $\mathcal{E}$ is linear and idempotent. In order to show that $\mathcal{E}$ is an orthogonal projection, we need to show that it is self-adjoint relative to the relevant inner product. Let us consider 
$\rho=\bigoplus \rho_\ell\otimes \tau_\ell$ and, as above:
\begin{eqnarray*}
X_\ell =&\Pi_{SF,\ell} X\,\Pi_{SF,\ell}=\sum_k A_{k,\ell}\otimes B_{k,\ell},\\
Y_\ell =&\Pi_{SF,\ell} Y\,\Pi_{SF,\ell}=\sum_j C_{j,\ell}\otimes D_{j,\ell}.
\end{eqnarray*}
If we apply Lemma \ref{lemma:blocks} to the operator 
$$W = \Ei_{\Ai_\rho}(X)\rho^{-1} = \bigoplus_\ell ( [\trace_{F,\ell}(X_\ell)\otimes\tau_\ell](\rho_\ell^{-1}\otimes\tau_\ell^{-1})),$$
we obtain:
\begin{eqnarray*}
\langle\mathcal{E}(X),Y \rangle_{\xi}&\hspace{-3mm}=&\hspace{-3mm}\trace(\Ei_{\Ai_\rho}(X)\rho^{-1}Y) \\
&\hspace{-3mm}=& \hspace{-3mm}\trace(\bigoplus_\ell \trace_{F,\ell}(X_\ell)\otimes \tau_\ell (\rho_\ell^{-1}\otimes \tau_\ell^{-1})Y_\ell)\\
&\hspace{-3mm}=& \hspace{-3mm}\sum_{\ell,k,j}\trace([A_{k,\ell}\trace(B_{k,\ell})\rho_\ell^{-1}\otimes I][C_{j,\ell}\otimes D_{j,\ell}])\\
&\hspace{-3mm}=&\hspace{-3mm}\sum_{\ell,k,j}\trace(B_{k,\ell})\trace(A_{k,\ell}\rho_\ell^{-1}C_{j,\ell})\trace( D_{j,\ell}).
\end{eqnarray*}
By similar calculation, 
\begin{eqnarray*}
\langle X,\mathcal{E}(Y)\rangle_{\xi} &=& \trace(X\rho^{-1}\mathcal{E}(Y)) \\
&=& \sum_{\ell,k,j}\trace(B_{k,\ell})\trace(A_{k,\ell}\rho_\ell^{-1}C_{j,\ell})\trace(D_{j,\ell}). 
\end{eqnarray*}
By comparison, we infer that $\langle\mathcal{E}(X),Y\rangle_\xi=\langle X,\mathcal{E}(Y)\rangle_\xi.$ 
A similar proof can be carried over using the  symmetric $\xi$-inner product of Eq. (\ref{eq:md_sym}).\qed}

We are now ready to prove the main results of this section. The first shows that the set of states with support on a 
target subspace can be made GAS by sequences of CPTP projections on larger subspaces that have the target as intersection.

{\thm [Subspace {stabilization}] 
\label{thm:alternating_pure} 
Let $\Hi_j$, $j=1,\ldots,r,$ be subspaces such that $\bigcap_j\Hi_j \equiv \hat\Hi.$ Then there exists CPTP projections 
${\mathcal{E}}_1,\ldots,{\mathcal{E}}_r$  onto ${\cal B}(\Hi_j)$, $j=1,\ldots,r,$ such that {
$\forall \tau \in \cal{D}(\hilbert)$: }
\beq
\label{eq:cptphalperin}
\lim_{n\rightarrow\infty}({\mathcal{E}}_r\ldots{\mathcal{E}}_1)^n 
(\tau)={\mathcal{E}}_{{\cal B}(\hat\Hi)}(\tau),
\eeq 
where ${\mathcal{E}}_{{\cal B}(\hat \Hi)}$ is a CPTP projection onto ${{\cal B}(\hat \Hi)}.$}

\proof 
We shall explicitly construct CPTP maps whose cyclic application ensures stabilization. 
Define $P_{j} $ to be the projector onto $\Hi_j$, and the CPTP maps:
\begin{equation}
\mathcal{E}_{j}(\cdot)  \equiv P_{j} (\cdot ) P_{j} + \frac{ P_{j} }{ \Tr ({P_{j}}) }  \,
\Tr \,(P^\perp_{j}\cdot ).  
\label{eq:newmaps}
\end{equation}
Consider $\hat P$ the orthogonal projection onto $\hat \Hi$ and the 
positive-semidefinite function
$V(\tau)=1-\Tr{}({\hat P \, \tau}),$ $\tau \in {\mathcal B}(\Hi).$
The variation of $V,$  when a $\Ei_j$ is applied, is
\[\Delta V (\tau)  {\equiv V(\Ei_j(\tau)) - V(\tau)} 
=-\Tr{} [{\hat P \, (\Ei_j(\tau)-\tau)}]\equiv \Delta V_j(\tau) .\]
If we show that this function is non-increasing along the trajectories generated by repetitions 
of the cycle of {all maps, namely, $\Ei_{\rm cycle}\equiv \Ei_r\circ \ldots \circ\Ei_1$}, 
the system is periodic thus its stability can be studied as a time-invariant one. Hence, by LaSalle-Krasowskii 
theorem \cite{lasalle-discrete}, the trajectories (being all bounded) will converge to the largest invariant set 
contained in the set of $\tau$ such that on a cycle $\Delta V_{\rm cycle}(\tau)=0$. We next show that this set 
must have support {\em only} on $\hat\Hi$. If an operator $\rho$ has support on $\hat\Hi,$ it is clearly invariant and 
$\Delta V(\rho)=0$.
Assume now that $\textup{supp}(\tau)\nsubseteq \Hi_j$ for some $j$, that is,  $\Tr{}{ (\tau P_j^\perp )} >0$.
By using the form of the map 
$\Ei_j$ given in Eq. (\ref{eq:newmaps}), we have 
\begin{eqnarray*}
\Delta V_j(\tau) 
&=&-\Tr{}{(\hat P (P_j \tau P_j ) )} - \Tr{}{(\tau P_j^\perp)} 
\frac{ \Tr{}{(\hat P (P_j))} }{  \tr{}{P_j }  }\nonumber\\
&+ &  \hspace*{3mm}  \Tr{}{(\hat P\tau)}
\label{eq:third}
\end{eqnarray*}
The sum of the first and the third term in the above equation 
is zero since $\hat P\leq P_j,$. The second term, on the other hand, is {\em strictly negative}. This is because: (i) we assumed that 
$\Tr ( {\tau P_j^\perp  } ) >0$;  (ii) with $\hat P\leq P_j,$ and ${\cal E}_j (P_j)$ 
having the same support of $P_j $ by construction, it also follows that
{$\tr{}{\Pi {\cal E}_j (P_j )}>0.$} This implies that $\Ei_j$ either leaves $\tau$ (and hence $V(\tau)$) invariant, or $\Delta V_j(\rho)<0.$ Hence, each cycle $\Ei_{\rm cycle}$
is such that $\Delta V_{\rm cycle}(\tau)=\sum_{j=1}^r\Delta V_j(\tau)<0$ for all $\tau\notin {\frak D}(\hat\Hi).$
We thus showed that no state $\tau$ with support outside of $\hat \Hi$ can be in the attractive set for the dynamics. 
Hence, the dynamics asymptotically converges onto ${\frak D}(\hat \Hi)$ which is the only invariant set for all the $\Ei_j$. 
\qed

The second result shows that the a similar property holds for more general fixed-point sets, as long as they 
contain a full-rank state:

{\thm [Full-rank fixed-set {stabilization}] 
\label{thm:quantum_map} 
Let the maps ${\mathcal{E}}_1,\ldots, {\mathcal{E}}_r$ be CPTP projections onto $\Ai_i$, $i=1,\ldots,r,$
and assume that $\hat\Ai \equiv \bigcap_{i=1}^r \Ai_i$ contains a full-rank state $\rho.$ Then 
{$\forall \tau \in \mathcal{D}(\Hi)$:}
\beq
\lim_{n\rightarrow\infty}({\mathcal{E}}_r\ldots{\mathcal{E}}_1)^n(\tau)={\mathcal{E}}_{\hat\Ai}(\tau),
\eeq 
where ${\mathcal{E}_{\hat\Ai}}$ is the CPTP projection onto $\hat\Ai.$}

\proof Let us consider $\xi=\rho^{-1}$; then $\rho \in \hat{\cal A}$ implies that the maps $\hat{\mathcal{E}}_i$ are all orthogonal projections with respect to the {\em same} $\rho^{-1}$-modified inner product (Propositions \ref{prop:proj}, \ref{prop:orth_proj_mod_in}). Hence, it suffice to apply von Neumann-Halperin, Theorem \ref{thm:halp}:
asymptotically, the cyclic application of orthogonal projections onto subsets converges to the projection onto 
the intersection of the subsets; in our case, the latter is $\hat\Ai$. \qed

\noindent Together with Theorem \ref{thm:modinvariance}, the above result implies that the intersection of 
fixed-point sets is still a fixed-point set of {\em some} map, as long as it contains a full-rank state:

{\cor 
If $\Ai_{i}$, $i=1,\ldots,r,$ are $\rho$-distorted algebras, with $\rho$ full rank, and are invariant for 
${\cal M}_{\rho,\frac{1}{2}},$ then $\hat\Ai=\bigcap_{i=1}^r \Ai_i$ is also a $\rho$-distorted algebra, 
invariant for ${\cal M}_{\rho,\frac{1}{2}}.$
}

\proof $\hat\Ai$ contains $\rho$ and the previous Theorem ensures that a CPTP projection onto it exists. Then by Theorem \ref{thm:modinvariance} it is invariant for ${\cal M}_{\rho,\frac{1}{2}}.$
\qed


Lastly, combining the ideas of the proof of Theorem \ref{thm:alternating_pure} and \ref{thm:quantum_map}, we obtain 
{\em sufficient} conditions for general fixed-point sets (that is, neither full algebras on a subspace nor containing a full-rank 
fixed-state).

{\thm [General fixed-point set {stabilization}]
\label{thm:alternating_general} 
Assume that the CPTP fixed-point sets $\Ai_i$, $i=1,\ldots,r,$ are such that $\hat\Ai \equiv 
\bigcap_{i=1}^r \Ai_i$ satisfies  
\[\supp(\hat\Ai)=\bigcap_{i=1}^r \supp(\Ai_i).\]
Then there exist CPTP projections ${\mathcal{E}}_1,\ldots,{\mathcal{E}}_r$ onto $\Ai_i$, $i=1,\ldots,r,$ 
such that {$\forall \tau \in \mathcal{D}(\Hi)$:}
\beq
\label{eq:cptphalperin1}
\lim_{n\rightarrow\infty}({\mathcal{E}}_r\ldots{\mathcal{E}}_1)^n(\tau)={\mathcal{E}}_{\hat\Ai}(\tau),\eeq 
where ${\mathcal{E}}$ is a CPTP projection onto $\hat\Ai.$}

\proof
To prove the claim, we explicitly construct the maps combining the ideas from the two previous theorems. 
Define $P_{j} $ to be the projector onto $\supp(\Ai_j)$, and the maps
\begin{eqnarray*}
\mathcal{E}^0_{j}(\cdot)  &\equiv& P_{j} (\cdot ) P_{j} + \frac{ P_{j} }{ \Tr ({P_{j}}) }  \,
\Tr \,(P^\perp_{j}\cdot ), \\
\mathcal{E}^1_{j} & \equiv &\mathcal{E}_{\Ai_j}\oplus {{\mathcal I}_{\Ai_j^\perp}, } 
\end{eqnarray*}
where $\mathcal{E}_{\Ai_j}:\mathcal{B}(\supp(\Ai_j))\to \mathcal{B}(\supp(\Ai_j))$ is the 
unique CPTP projection onto $\Ai_j$ (notice that on its own support $\Ai_j$ includes a full-rank state), and 
${\mathcal I}_{\Ai_j^\perp}$ denotes the identity map on operators on $\supp(\Ai_j)^\perp$.
Now construct
\[\Ei_j (\cdot)\equiv \mathcal{E}^1_{j}\circ \mathcal{E}^0_{j}(\cdot).\]
Since {each map $\mathcal{E}^1_{j}$ leaves the support of $P_j$ invariant,}
the same Lyapunov argument of Theorem \ref{thm:alternating_pure} 
shows that:
\beq
\label{eq:par1}
\supp(\lim_{n\rightarrow\infty}({\mathcal{E}}_r\ldots{\mathcal{E}}_1)^n(\rho))\subseteq\supp({\mathcal{E}}_{{\cal B}(\hat\Hi)}(\tau)).
\eeq
We thus have that the largest invariant set for a cycle of maps ${\mathcal{E}}_r\ldots{\mathcal{E}}_1$
has support equal to $\hat\Ai,$ and by the discrete-time invariance principle \cite{lasalle-discrete}, 
the dynamics converge to that.

Now notice that, since $\hat\Ai$ is contained in each of the $\Ai_j=\fix(\Ei_j)$, such is any maximum-rank operator in $\hat\Ai,$ which implies (see e.g. Lemma 1 in \cite{cirillo-decompositions}) that $\supp(\hat\Ai)$ is an invariant subspace for each $\Ei_j.$ Hence, $\Ei_j$ restricted to ${\cal B}(\supp(\hat\Ai))$ is still CPTP, and by construction projects onto the elements of $\Ai_j$ that have support contained in $\supp(\hat\Ai).$ Such a set, call it $\hat\Ai_j$, is thus a valid fixed-point set. By Theorem \ref{thm:quantum_map}, we have that on the support of $\hat\Ai$ the limit in Eq. \eqref{eq:cptphalperin1} converges to $\hat\Ai.$ This shows that the largest invariant set for the cycle is exactly $\hat\Ai,$ hence the claim 
is proved.
\qed

\noindent 
{\em Remark:} In order for the proposed quantum alternating projection methods to be effective, 
it is important that the relevant CPTP maps be sufficiently simple to evaluate and implement.  
Assuming that the map $\Ei$ is easily achievable, it is useful to note that the projection 
map $\Ei_{{\cal A}_\rho}$ defined in Eq. (\ref{einfty}) may be approximated through iteration of a map 
$\tilde{\Ei}_\lambda \equiv (1-\lambda) \Ei + \lambda {\cal I}$, where $\lambda \in (0,1)$. 
Since $\tilde{\Ei}_\lambda$ has 1 as the only eigenvalue on the unit circle \cite{mazzarella-qconsensus}, it is easy to show that
$ \lim_{n \rightarrow \infty} \tilde{\Ei}_\lambda^n = \Ei_{{\cal A}_\rho}$, $\Ei_{{\cal A}_\rho} \approx 
\tilde{\Ei}_\lambda^n$ for a sufficiently large number of iterations.

\subsection{Convergence rate} 
\label{sec:speed}

For practical applications, a relevant aspect of stabilizing a 
target set is provided by the rate of asymptotic convergence.  

In our case, focusing for concreteness on state stabilization, 
the key to Theorems \ref{thm:quantum_map} and \ref{thm:alternating_general} (and to  
Theorem \ref{thm:mixedqls} that will be given in Sec. \ref{sec:qls}) is Halperin's alternating projection result. 
Thus, if we are interested in the speed of convergence of stabilizing dynamics for a {\em full-rank state} $\rho$, 
this is just the speed of convergence of the classical alternating projection method.  As mentioned, and as 
we recall in Appendix \ref{app:cosine}, the rate is related to the {angles} between the subspaces. In fact, 
an upper bound for the distance decrease from the target attained in $n$ repetitions of a cycle of maps 
$\Ei_{\rm cycle}= \Ei_r\circ \ldots \circ\Ei_1$ is 
given in terms of the {\em contraction coefficient} 
$c^{\frac{n}{2}},$ where $c \equiv 1-\prod_{i=1}^{r-1} \sin^2{\theta_i},$ and $\theta_i$ 
is the angle between $\Ai_i$ and the intersection of the fixed points of the following maps.
In particular, by Theorem \ref{thm:conv_n} in the Appendix, it follows that a sufficient condition for 
{\em finite-time convergence} of iterated projections is given by $c=0,$ which is satisfied for example if $c(\mathcal{A}_i,\mathcal{A}_j)=0$ for all $1\leq i,j\leq r$. That is, equivalently, 
$$\big [\mathcal{A}_i\cap\big(\bigcap_{t=1}^r\mathcal{A}_t\big)^\perp\big ]\perp_\rho \big[\mathcal{A}_j\cap\big(\bigcap_{t=1}^r\mathcal{A}_t\big)^\perp\big],$$ for every $i,j=i+1,\ldots,r$, where orthogonality is with respect to the $\rho^{-1}$-inner product, either symmetric or not. However, this condition is clearly {\em not} necessary.

For a {\em pure target state} $\rho$, a natural way to quantify the convergence rate is to consider the decrease of a 
suitable Lyapunov function. Given the form of the projection maps we propose, a natural choice is 
the same $V$ we use in the proof of Theorem \ref{thm:alternating_pure}, namely, 
$V(\tau) = 1-\Tr{}({\rho \, \tau}).$
The variation of $V,$  when $\Ei_{\rm cycle}$ is applied, is
\[\Delta V(\tau)=-\Tr{} [{\rho(\Ei_{\rm cycle}(\tau)-\tau)}] <0, \quad \forall \tau . \]
\noindent 
The contraction coefficient in the pseudo-distance $V$ is then:
\[c=\max_{\tau\geq 0,\Tr(\tau)=1, \Tr({\tau\rho)=0} }
\Delta V(\tau).\]
In this way, we select the rate corresponding to the worst-case state with support orthogonal to the target (notice that maximization over all states would have just given zero).

\subsection{Robustness with respect to randomization}
\label{sec:robustness}

While Theorem \ref{thm:alternating_pure} and Theorem \ref{thm:quantum_map} require {\em deterministic} cyclic repetition of the CPTP projections, the order is not critical for convergence. Randomizing the order of the maps still leads to asymptotic convergence, 
albeit in probability. 
We say that an operator-valued process $X(t)$ {\em converges in probability} to $X^*$ if, for any 
$\delta,\varepsilon > 0$, there exists a time $T>0$ such that
$$\mathbb{P}[\, \trace((X(T)-X_*)^2) \,  > \varepsilon \,] \; < \delta \, .$$
Likewise, $X(t)$ {\em converges in expectation} if $\mathbb{E}(\rho(t))\to\rho^*$ when $t\to +\infty.$
Establishing convergence in probability uses the following lemma, adapted from 
\cite{ticozzi-newqconsensus}, a consequence of the second Borel-Cantelli lemma:
\begin{lemma}[Convergence in probability]
\label{lemma:rand}
Consider a finite number of CPTP maps $\{\Ei_j\}_{j=1}^M,$ and a (Lyapunov) function $V(\rho),$ such that 
$V(\rho)\geq 0$ and $V(\rho)=0$ if and only if $\rho\in{\cal S},$ with $\cal S\subset {\mathcal D}(\Hi)$ 
some set of density operators. Assume, furthermore that:
\begin{itemize}
\item[(i)] For each $j$ and state $\rho$, $V(\Ei_j(\rho))\leq V(\rho)$.
\item[(ii)] For each $\varepsilon> 0$ there exists a finite sequence of maps 
\beq
\label{eq:contrseq}
\Ei_{\varepsilon}=\Ei_{j_K}\circ \ldots\circ \Ei_{{j_1}},
\eeq
with $j_\ell\in\{1,\ldots, M \}$ for all $\ell$, such that 
$V(\Ei_{\varepsilon}(\rho))< \varepsilon$ for all $\rho\neq{\cal S}.$
\end{itemize}
{Assume that the maps are selected at random, with independent probability distribution $\P_t[\Ei_{j}]$ 
at each time $t$, and that
there exists $\varepsilon>0$ for which $\P_t[\Ei_{j}] > \varepsilon$ for all $t$. 
Then, for any $\gamma>0$, the probability of having $V(\rho(t))<\gamma$ converges to $1$ as $t\rightarrow +\infty$.}
\end{lemma}

\vspace*{1mm}

Using the above result, we can prove the following:
\begin{cor} 
Let ${\mathcal{E}}_1,\ldots,{\mathcal{E}}_r$ CPTP projections onto $\Ai_i={\mathfrak B}(\Hi_i)$, $i=1,\ldots, r$.
Assume that at each step $t\geq 0$ the map $\Ei_{j(t)}$ is selected  randomly from a probability distribution 
\[ \Big\{ q_j(t)=\mathbb{P}[{{\cal E}_{j(t)}}] > 0 \vert \sum_{j} q_{j}(t) = 1 \Big\},\] 
and that $q_j(t)>\epsilon>0$ for all $j$ and $t\geq 0.$ For all $ \tau \in \mathfrak{D}(\hilbert)$, let $\tau(t) \equiv 
\Ei_{j(t)}\circ\ldots \circ\Ei_{j(1)}(\tau).$ Then $\tau(t)$ converges {in probability and in expectation} to 
$$
\tau^*={\mathcal{E}}_{\hat\Ai}(\tau),
$$ 
where ${\mathcal{E}}$ is the CPTP projection onto $\hat\Ai.$
\end{cor}

\proof 
Given Lemma \ref{lemma:rand}, it suffices to consider \(V(\tau)\equiv 1-\rm{Tr}(\hat P \, \tau).\) It is non-increasing, and Theorem \ref{thm:quantum_map} also ensures that for every $\varepsilon>0$, there exists a finite number of cycles of the maps that makes $V(\tau)< \varepsilon.$
\qed

A similar result holds for the full-rank case:

\begin{cor} 
Let ${\mathcal{E}}_1,\ldots,{\mathcal{E}}_r$ CPTP projections onto $\Ai_i$, $i=1,\ldots,r,$ and assume that 
$\hat\Ai=\bigcap_{i=1}^r \Ai_i$ contains a full-rank state $\rho.$ Assume that at each step $t\geq 0$ the map $\Ei_{j(t)}$ is selected  randomly from a probability distribution 
\[\Big \{ q_j(t)=\mathbb{P}[{{\cal E}_{j(t)}}] > 0 \vert \sum_{j} q_{j}(t) = 1 \Big\},\] 
and that $q_j(t)>\epsilon>0$ for all $j$ and $t\geq 0.$ For all $ \tau \in \mathfrak{D}(\hilbert)$, let $\tau(t) \equiv\Ei_j(t)\circ\ldots \circ\Ei_j(1)(\tau).$ Then $\tau(t)$ converges {in probability and in expectation} to 
$$ 
\tau^*={\mathcal{E}}_{\hat\Ai}(\tau),
$$  
where ${\mathcal{E}}$ is the CPTP projection onto $\hat\Ai.$
\end{cor}

\proof 
Given the Lemma \ref{lemma:rand}, it suffices to consider \(V(\tau)\equiv \langle (\tau-\tau^*),(\tau-\tau^*)\rangle_{\rho^{-1}}.\) It is non-increasing, and  Theorem \ref{thm:quantum_map} ensures that  for every $\varepsilon>0$ there exists a finite number of cycles of the maps that makes $V(\tau)< \varepsilon.$
\qed

\section{Quasi-Local State Stabilization}\label{sec:ql}

\subsection{Locality notion}

In this section we specialize to a multipartite quantum system
consisting of $n$ (distinguishable) subsystems, or ``qudits'', defined on a tensor-product 
Hilbert space
$$\Hi \equiv \bigotimes_{a=1}^n\Hi_a, \quad a=1,\ldots,n,\;
\text{dim}(\Hi_a)=d_a, \, \text{dim}(\Hi)=d. $$
In order to impose {\em quasi-locality constraints} on operators 
and dynamics on $\Hi,$ we introduce {\em neighborhoods}. Following
\cite{ticozzi-ql,ticozzi-ql1,johnson-ffqls}, neighborhoods $\{ {\cal N}_j \}$ are subsets 
of indexes labeling the subsystems, that is,
\[{\cal N}_j\subsetneq\{1,\ldots,n\}, \quad 
{j=1,\ldots, K.}\]
A {\em neighborhood operator} $M$ is an operator on $\Hi$ such that there exists a 
neighborhood ${\cal N}_j$ for which we may write 
\[ M \equiv M_{{\cal N}_j}\otimes I_{\overline{\cal N}_j},\] 
where $M_{{\cal N}_j}$ accounts for the action of $M$ on subsystems in 
${\cal N}_j$, and $I_{\overline{\cal N}_j}\equiv \bigotimes_{a\notin{\cal N}_j}I_a$ is 
the identity on the remaining ones.
Once a state $\rho \in {\cal D}(\Hi)$ and a neighborhood structure are assigned on $\Hi,$ {\em
reduced neighborhood states} may be computed via partial trace as usual: 
\begin{equation} 
\label{redstate} 
\rho_{{\cal N}_j} \equiv \mbox{Tr}_{\overline{\cal N}_j}(\rho),
\quad \rho \in {\mathfrak D}(\Hi), \;\; j=1,\ldots, K,
\end{equation}
where $\trace_{\overline{\cal N}_j}$ indicates the partial trace over the 
tensor complement of the neighborhood ${\cal N}_j,$ namely,
$\Hi_{\overline{\cal N}_j} \equiv \bigotimes_{a\notin{\cal N}_j}\Hi_a$.  
{A strictly ``local'' setting corresponds 
to the case 
where ${\cal N}_j \equiv \{ j\}$, that is, each subsystem forms a distinct neighborhood.} 

Assume that some quasi-locality notion is {\em fixed} by specifying a set of neighborhoods, 
{${\cal N}\equiv \{ {\cal N}_j\}$.}
A {\em CP map $\Ei$ is a neighborhood map} relative to  ${\cal N}$ if, for some $j$, 
\begin{equation}
\Ei=\Ei_{{\cal N}_j}\otimes {\cal I}_{\overline{\cal N}_j},\
\label{eq:neigh}
\end{equation}
where $\Ei_{{\cal N}_j}$ is the restriction of $\Ei$ to operators on the subsystems in ${\cal N}_j$ and 
${\cal I}_{\overline{\cal N}_j}$ is the identity map for operators on $\Hi_{\overline{\cal N}_j}$. 
An equivalent formulation can be given in terms of the OSR: that is, 
$\Ei(\rho)=\sum_k M_k \rho M_k^\dag$ is a neighborhood map relative to  ${\cal N}$
if there exists a neighborhood
${\cal N}_j$ such that, {\em for all} $k,$ \[M_k = M_{{\cal N}_j,k}\otimes I_{\overline{\cal N}_j}.\] 
The reduced map on the neighborhood is then \[\Ei_{{\cal N}_j}(\cdot )=\sum_k M_{{\cal N}_j,k}\,
\cdot\,M_{{\cal N}_j,k}^\dag.\]
Since the identity factor is preserved by sums (and products) of the $M_k$, it is immediate to verify that the 
property of {$\Ei$ being a neighborhood map} is well-defined with respect to the freedom in the OSR \cite{nielsen-chuang}.  

\begin{defin} 
(i) A state $\rho$ is  discrete-time {\em Quasi-Locally Stabilizable} (QLS) if there exists a sequence $\{\Ei_t\}_{t\geq 0}$ of neighborhood  maps such that $\rho$ is GAS for the associated propagator $\Ei_{t,s}=\Ei_{t-1}\circ\ldots\circ\Ei_s$, 
namely:
\begin{eqnarray}
&&
\hspace*{-5mm} \Ei_{t,s}(\rho)=\rho,\quad\forall t\geq s\geq 0;
\label{pinvQLS}\\
&&
\hspace*{-5mm}
\lim_{t\rightarrow\infty} \|\Ei_{t,s}(\sigma),\rho\|=0,\quad \forall \sigma\in\DD,\, \forall s\geq 0.
\label{pattQLS}
\end{eqnarray} 
(ii) The state is {\em QLS in finite time} (or finite-time QLS) 
if there exists a finite sequence of $T$ maps whose propagator 
satisfies the invariance requirement of Eq. \eqref{pinvQLS} and
\beqa 
{\Ei_{T,0}(\sigma)=\rho, \quad \forall \sigma\in\DD.}
\label{FTQLS}
\eeqa 
\end{defin}

\noindent{\em Remark:} 
With respect to the definition of quasi-locality that naturally emerges for continuous-time Markov dynamics 
\cite{ticozzi-ql,ticozzi-ql1,johnson-ffqls}, it is important to appreciate that constraining discrete-time dynamics 
to be QL in the above sense is more restrictive. 
In fact, even if a generator ${\cal L}$ of a continuous-time (homogeneous)
semigroup  can be written as a sum of 
neighborhood generators,  namely, ${\cal L}=\sum_k {\cal L}_k$, 
the generated semigroup $\Ei_t \equiv e^{\Li t}, t\geq 0,$  is {\em not}, in general, QL in the sense of 
Eq. (\ref{eq:neigh}) at any time. In some sense, one may think of the different noise components $\Li_1,\ldots,\Li_k$ 
of the continuous-time generator as acting ``in parallel''. On the other hand, were the maps $\Ei_j$ we consider in 
this paper each generated by some corresponding neighborhood generator ${\Li_j},$ then 
by QL discrete-time dynamics we would be requesting that, on each time interval, a {\em single} noise 
operator is active, thus obtaining global {switching}  dynamics \cite{scaramuzza-switching} of the form
\[ e^{\Li_k T_k}\circ e^{\Li_{k-1} T_{k-1}}\circ\ldots\circ e^{\Li_1 T_1}.\]

We could have requested each $\Ei_t$ to be a convex combination of neighborhood maps 
acting on different neighborhoods, however it is not difficult to see that this case can be studied as 
the convergence in expectation for a randomized sequence.
Hence, we are focusing on the {\em most restrictive definition of QL constraint} for discrete-time Markov 
dynamics. With respect to the continuous dynamics, however, we allow for the evolution to be 
time-inhomogeneous. Remarkably, we {shall find a characterization of QLS pure states 
that is equivalent to the continuous-time case, when the latter dynamics} are required to be
 {\em frustration free} (see Section \ref{sec:physint}).

\subsection{Invariance conditions and minimal fixed point sets}
\label{sec:invariance}

In this section, we build on the invariance requirement of Eq. \eqref{pinvQLS} to find {\em necessary} conditions that the discrete-time dynamics must satisfy in order to have a given state $\rho$ as its unique and attracting equilibrium. These impose a certain minimal fixed-point set, and hence suggest a structure for the stabilizing dynamics.

Following \cite{johnson-ffqls}, given an operator $X\in \mathcal{B}(\hilbert_A\otimes \hilbert_B)$, with corresponding (operator) Schmidt decomposition $X=\sum_j A_j\otimes B_j,$
we define its {\em Schmidt span} as: 
$$\Sigma_A(X) \equiv \span(\{A_j\}).$$
\noindent
The Schmidt span is important because, if we want to leave an operator invariant with a neighborhood map, 
this also imposes the invariance of its Schmidt span. The following lemma, proved in \cite{johnson-ffqls}, makes 
this precise:
\begin{lemma}
\label{invSS}
{ 
Given a vector $v\in V_{A}\otimes V_{B}$ and $M_A\in\mathcal{B}(V_A)$, if $(M_A\otimes \identity_B) v=\lambda v$, then $(M_A\otimes\identity_B) v'=\lambda v'$ for all $v'\in\Sigma_{A}(v)\otimes V_B$.}
\label{thm:localkernel}
\end{lemma}

What we need here can be obtained by adapting this result to our case, specifically:
{\begin{cor} 
\label{cor:fixed}
Given a $\rho \in \mathcal{D}(\Hi_{\mathcal{N}_j}\otimes \hilbert_{\overline{{\mathcal N}}_j})$ and 
a neighborhood $\mathcal{E}=\mathcal{E}_{\mathcal{N}_j}\otimes I_{\overline{\mathcal{N}}_j}$, then
$\span(\rho)\subseteq \fix(\mathcal{E})$ implies
$$\Sigma_{\mathcal{N}_j}(\rho)\otimes \mathcal{B}(\hilbert_{\overline{\mathcal{N}}_j})
\subseteq \fix(\mathcal{E}_{{\cal N}_j}).$$
\end{cor}}

A Schmidt span need {\em not} be a valid fixed-point set, namely, a $\rho$-distorted algebra 
that is invariant for ${\cal M}_{\rho,\frac{1}{2}}.$ In general, we need to further enlarge the QL fixed-point sets from the Schmidt span to suitable algebras.  We discuss separately two relevant cases.

$\bullet$ {\em Pure states.---} Let $\rho=\ket{\psi}\bra{\psi}$ be a pure state and assume that, with respect to the factorization $\Hi_{\Ni_j}\otimes\Hi_{\overline\Ni_j},$ its Schmidt decomposition $\ket{\psi}=\sum_k c_k\ket{\psi_k}\otimes\ket{\phi_k}.$ Let {
$\Hi^0_{\Ni_j} \equiv \span\{\ket{\psi_k} \}  = \text{supp} (\rho_{\Ni_j} )$. }
Then we have \cite{johnson-ffqls}:
\beq
\label{eq:purecase}
\Sigma_{\Ni_j}(\rho)={\cal B}(\Hi^0_{\Ni_j}).
\eeq
In this case the Schmidt span is indeed a valid fixed-point set, and no further enlargement is needed. The {\em minimal 
fixed-point set} 
for neighborhood maps required to preserve 
$\rho$ is thus 
$\Fi_j \equiv {\cal B}(\Hi^0_{\Ni_j}) \otimes \mathcal{B}(\hilbert_{\overline{\mathcal{N}}_j})  .$ 
By construction, each $\Fi_j$ contains $\rho.$ Notice that their intersection is just $\rho$ if and only if
\beq
\label{eq:purecaseket} 
\span\{\ket{\psi}\}=\bigcap_j \Hi^0_{\Ni_j}  \otimes \Hi_{\overline{\Ni}_j}=\bigcap_j {\Hi^0_j,  }
\eeq 
where we have defined $\Hi^0_j\equiv \Hi^0_{\Ni_j}  \otimes \Hi_{\overline{\Ni}_j}.$

$\bullet$ {\em Full rank states.---} If $\rho$ is a full-rank state, and $W$ a set of operators, the {\em minimal fixed-point set generated by $\rho$ and $W$}, by Theorem \ref{thm:modinvariance}, is the smallest $\rho$-distorted algebra generated by $W$ which is invariant with respect to ${\cal M}_{\rho,\frac{1}{2}}$.
Notice that, since $\rho$ is full rank, its reduced states $\rho_{\mathcal{N}_j}$ are also full rank.
{
Denote by $\alg_\rho(W)$ the $\dag$-closed $\rho$-distorted algebra generated by $W.$ Call $W_j \equiv \Sigma_{\mathcal{N}_j}(\rho)$. The minimal fixed-point 
sets $\mathcal{F}_{\rho_{\mathcal{N}_j}}(W_j )$ can then be constructed iteratively from 
$\mathcal{F}_j^{(0)} \equiv \alg_{\rho_{\mathcal{N}_j}}(W_j)$, with the $k$-th step given by 
\cite{johnson-ffqls}:  
$$\mathcal{F}_j^{(k+1)} \equiv \alg_{\rho_{\mathcal{N}_j}} 
{({\cal M}_{  \rho_{\mathcal{N}_j}, \frac{1}{2}  }  (\Fi_j^{(k)}),{\Fi_j^{(k)}}).}$$
We keep iterating until $\mathcal{F}_j^{(k+1)}=\mathcal{F}_j^{(k)} = \mathcal{F}_{\rho_{{\cal N}_j}} (W_j).$
When that happens, define 
\begin{equation}
\mathcal{F}_j \equiv \mathcal{F}_{\rho_{{\cal N}_j}} ( \Sigma_{\mathcal{N}_j}(\rho) ) \otimes {\cal B}(\Hi_{\overline\Ni_j}).
\label{eq:mixedcase}
\end{equation}}
\noindent 
Since the ${\cal F}_j$ are constructed to be the minimal sets for neighborhood maps that contain the 
given state and its corresponding Schmidt span, then clearly:
\[\span(\rho)\subset\bigcap_j \mathcal{F}_j.\]

\subsection{Stabilizability under quasi-locality constraints}
\label{sec:qls}

In the case of a pure target state, we can prove the following:
\begin{thm}[QLS pure states]
\label{mainthm} 
A pure state $\rho=|\psi\rangle\langle \psi|$ is QLS by discrete-time dynamics if and only if 
\beq
\label{GAScond}
{ \supp(\rho)= \bigcap_j {{\mathcal H}^0_{j}.} }
\eeq
\end{thm}

\proof Given Corollary \ref{cor:fixed},
any dynamics that make $\rho$ QLS (and hence leaves it invariant) must consist of neighborhood maps $\{\Ei_j\}$ with corresponding fixed points such that:
\[\Fi_{k}\subseteq \fix(\Ei_j),\]
whenever $\Ei_j$ is a $\Ni_{k}$-neighborhood map. If the intersection of {the fixed-point sets} is not unique, then $\rho$ cannot be GAS, since there would be another state that is not attracted to it. Given Eq. \eqref{eq:purecaseket}, we have 
\[ {
\span(\rho)=\bigcap_k \mathcal{F}_k \quad \Longleftrightarrow \quad 
\supp(\rho)= \bigcap_j {{\mathcal H}^0_{j}.},   }\]
which proves necessity.
For sufficiency, we explicitly construct neighborhood maps whose cyclic application ensures stabilization. Define$P_{\neigh_j} $ to be the projector onto $\textup{supp}(\rho_{\neigh_j} )$, and the CPTP maps:
\begin{equation}
\mathcal{E}_{\neigh_j}(\cdot)  \equiv P_{\neigh_j} (\cdot ) P_{\neigh_j} + \frac{ P_{\neigh_j} }{ \Tr ({P_{\neigh_j}}) }  \,
\Tr \,(P^\perp_{\neigh_j}\cdot ),   
\label{newmaps}
\end{equation}
with $\Ei_j \equiv \mathcal{E}_{\neigh_j}\otimes {\mathcal I}_{\overline\neigh_j}.$
Consider the 
positive-semidefinite function
$V(\tau)=1-\tr{}{\rho \, \tau},$ $\tau \in {\mathcal B}(\Hi).$
The result then follows from Theorem \ref{thm:alternating_pure}. 
\qed

\noindent An equivalent characterization can be given for {full-rank target states}:

\begin{thm}[QLS full-rank states]
\label{thm:mixedqls}
A full-rank state $\rho\in {\cal D}(\Hi)$ is QLS by discrete-time dynamics if and only if 
\beq
\label{eq:algint}
\span(\rho)=\bigcap_k \mathcal{F}_k\eeq 
\end{thm}

{\proof As before, by contradiction, suppose that  $\rho_2\in \bigcap_k \mathcal{F}_k$ exists, such that $\rho_2\not = \rho$. This clearly implies that $\rho$ cannot be GAS because there would exist another invariant state, which is not attracted to $\rho$. This proves necessity.
Sufficiency derives from the alternating CPTP projection theorem.  Specifically, let $\mathcal{E}_{\Ni_k}$ be the CPTP projection onto ${\cal F}_k,$ and $$\Ei_k\equiv \mathcal{E}_{\Ni_k}\otimes {\rm Id}_{\overline{\Ni}_k}.$$ By Theorem \ref{thm:quantum_map}, we already know that for every $\rho$, $(\mathcal{E}_M\ldots \mathcal{E}_1)^k(\rho)\rightarrow \bigcap_k \mathcal{F}_k$ for $k\rightarrow \infty$. Now, by hypothesis, $\bigcap_k \mathcal{F}_k=\span(\rho)$ and, being $\rho$ the only (trace one) state in his own span,  $\rho$ is GAS. \qed 

\noindent A set of {\em sufficient conditions}, stemming from Theorem \ref{thm:alternating_general}, can be also derived in an analogous way 
for a general target state.

\subsection{Physical interpretation: discrete-time quasi-local stabilizability is equivalent to 
cooling without frustration}

\label{sec:physint}
 
Consider a {\em quasi-local Hamiltonian}, that is, $H=\sum_k H_k,$ 
$H_k=H_{{\cal N}_k}\otimes I_{\overline{\cal N}_k}.$ 
{$H$ is called a {\em parent} Hamiltonian for a pure state $\ket{\psi}$ 
if it admits $\ket{\psi}$ as a ground state, and it is 
called a is called a {\em frustration-free (FF)} Hamiltonian if any global 
ground state is {\em also} a local ground state \cite{perezgarcia2007}, that is, 
\[ \text{argmin}_{\ket{\psi} \in \hilbert} \bra{\psi} H \ket{\psi} \subseteq 
\text{argmin}_{\ket{\psi} \in \hilbert} \bra{\psi} H_k \ket{\psi}, \forall k. \] }
\noindent 
Suppose that a target state $\ket{\psi}$ admits a FF QL parent Hamiltonian $H$ for which it is 
the {\em unique} ground state. 
Then, similarly to what has been done for continuous-time dissipative preparation 
\cite{kraus-dissipative,ticozzi-ql}, the structure of $H$ may be naturally used to derive a
stabilizing discrete-time dynamics: it suffices to implement neighborhood maps $M_k$
that stabilize the eigenspace associated to the minimum eigenvalue of each $H_k$. These 
can thought as maps that locally ``cool'' the system. In view of 
{Theorem \ref{mainthm},} it is easy to show that this
condition is also necessary:

\begin{cor}
\label{parent}
{A state $\rho=|\psi\rangle\langle\psi|$ is QLS by discrete-time dynamics if and only if it is 
the unique ground state of a FF QL parent Hamiltonian.}
\end{cor}

\proof Without loss of generality we can consider {FF}  QL Hamiltonians
$H=\sum_k H_k,$ where each $H_k$ is a projection. Let $\rho$ satisfy
Eq. \eqref{GAScond}, which is equivalent to be QLS, and define $H_k \equiv 
\Pi^\perp_{{\cal N}_k}\otimes I_{\overline{\cal N}_k}$, with $\Pi^\perp_{{\cal N}_k}$ 
being the orthogonal projector
onto the {orthogonal complement} of the support of $\rho_{{\cal N}_k}$, that is,
$\Hi_{{\cal N}_k}\ominus \supp({\rho_{{\cal N}_k}})$. Given Theorem
\ref{mainthm}, $|\psi\rangle $ is the unique pure state in
$\bigcap_k\supp(\rho_{{\cal N}_k}\otimes I_{\overline{\cal N}_k}),$ and
thus the unique state in the kernel of all the $H_k$. Conversely, if a FF 
QL parent Hamiltonian exists, the kernels of the $H_k$ satisfy the QLS condition 
and to each $H_k$ we can associate a CPTP map as in \eqref{newmaps} 
that projects onto its kernel. \qed

An equivalent characterization works for generic, full-rank target states, 
but we need to move from Hamiltonians to semigroup generators, while maintaining 
frustration-freeness in a suitable sense. 
{If, as before, ${\cal E}_{t}=e^{{\cal L}}$ is  
the propagator arising from a time-invariant QL generator ${\cal L}$, }
we are interested in QL generators whose neighborhood components drive the system to a global equilibrium which is 
{\em also} a local equilibrium for each of them separately. That is, following \cite{Brandao2014,johnson-ffqls}:

\begin{defin} 
{A QL generator $\mathcal{L}=\sum_{j}\mathcal{L}_{j},
$ where ${\cal L}_j$ are neighborhood generators, is \emph{Frustration Free} (FF) relative to a neighborhood structure $\Ni =\{ \Ni_j\}$ if \[\rho\in\ker (\Li)\quad \iff \quad \rho \in \ker(\Li_j), \quad \forall j.\] }
\end{defin}

\noindent It is worth noting that a state which is invariant for all the local generators is always an equilibrium: the real requirement is that these states are {\em all} the equilibria. We then have:
\begin{prop}
\label{prop:cooling} 
A pure or full-rank state $\rho$ is discrete-time QLS if and only if it is QLS via FF continuous-time dynamics, that is, there exists a FF generator ${\cal L}$ with respect to the same neighborhood structure $\neigh$ such that
\[\lim_{t\to+\infty}e^{{\cal L}t}\rho_0=\rho,\quad \forall\rho_0.\]
\end{prop}
\proof The claim follows from Theorems 7 and 8 in \cite{johnson-ffqls}, which characterize the states that are unique fixed points of a FF generator as precisely the states that satisfy Eq. \eqref{eq:algint}. 
\qed

{\em Remark:} Based on the above results, the conditions that guarantee either a pure or a full-rank state 
to be QLS in discrete time are the same that guarantee existence of a FF stabilizing generator 
in continuous time.  We stress that if more general {continuous-time} generators are allowed, namely, 
frustration is permitted as in Hamiltonian-assisted stabilization \cite{ticozzi-ql1}, then the continuous-time setting 
can be more powerful. On the one hand, considering the stricter nature of the QL constraint for the discrete-time 
setting, this is not surprising. On the other hand, if Liouvillian is no longer FF, then the target is globally invariant 
for ${\cal L}$ but no longer invariant for individual QL components ${\cal L}_j$, suggesting that a weaker 
(``stroboscopic'') invariance requirement could be more appropriate to ``mimic'' the effect of frustration 
in the discrete-time QL setting.

\subsection{Classes of discrete-time QLS states}

Being the conditions for discrete-time QLS states the same as in continuous time with FF dynamics, we may 
conclude whether certain classes of states are QLS or not by exploiting the results already established in 
\cite{ticozzi-ql,ticozzi-ql1,johnson-ffqls}. While we refer to the original references for additional detail 
and context, some notable examples are summarized in what follows.
Among {\em pure states}:
\begin{enumerate}
\item {\em Graph states} \cite{graphstates} (more generally, {\em stabilizer states} \cite{nielsen-chuang} that 
admit stabilizer group generators that are neighborhood operators) are discrete-time QLS. 
These states are entangled, and are a key resource for one-way quantum computation.

\item Certain, but not all, {\em Dicke states} are discrete-time QLS. Dicke states are symmetric with respect 
to subsystem permutations, and have a specified ``excitation number'' \cite{dicke}.  Dicke states exhibit 
entanglement properties that are, in some sense, robust: some entanglement is preserved even if some subsystems are measured or traced out. 
The $n$-qubit single-excitation Dicke state, also known as W-state,   
$$\ket{\psi^n_{\text{W}}} \equiv \frac{1}{\sqrt{n}} (\ket{100\ldots 0}+\ket{010\ldots 0}+\ldots
+\ket{000\ldots 1}), $$ 
fails to satisfy the conditions of Theorem \ref{mainthm}, so $\rho_{\text W}$ is {\em not} discrete-time QLS for non-trivial neighborhood structures (that is, unless there is a neighborhood that covers the whole network). On the other hand, for example, the two-excitation Dicke state on $n=4$ qubits, 
\beqan
&&\hspace{-8mm}\ket{\psi_D^4}\equiv \\&&\hspace{-8mm}\frac{\ket{1100}+\ket{1010}+\ket{1001}+\ket{0110}+
\ket{0101}+\ket{0011}}{\sqrt{6}},
\eeqan is QLS. A more general class of QLS Dicke states on qudits is presented in \cite{johnson-ffqls}.
\noindent 
\end{enumerate}
Among {\em full-rank states}:
\begin{enumerate}
\item {\em Commuting Gibbs states} are discrete-time QLS with respect to a suitable locality notion 
\cite{johnson-ffqls}. A Gibbs state represent the canonical
thermal equilibrium state for a statistical system at temperature $\beta^{-1}$: if a chain of qudits 
is associated to a nearest-neighbor (NN) Hamiltonian $H=\sum_k H_k$, its Gibbs state is 
\[\rho_\beta \equiv \frac{e^{-\beta H}}{\trace(e^{-\beta H})}.\]
If the Hamiltonian is commuting, namely, $[H_j,H_k]=0$ for all $j,k,$ then $\rho$ is QLS with respect to 
an {\em enlarged} QL notion, where ${\cal N}^0_j$ contains all subsystems that belong to a NN neighborhood ${\cal N}_k$ such that ${\cal N}_k\cap {\cal N}_j\neq\emptyset.$ In analogy to the continuous-time case \cite{Brandao2014}, this shows that Gibbs samplers  based on QL discrete-time dissipative dynamics 
are also viable, at least in the commuting case.


\item {\em Certain mixtures of factorized and entangled states} are discrete-time QLS. For example, consider a $4$-qubits system and the family of states parametrized by $\epsilon \in (0,1)$:
\[ \rho_\epsilon \equiv (1-\epsilon) \, \ketbra{\psi_D^4}
+ \epsilon \, \ketbra{\textup{GHZ}^4}, \] 
where $\ket{\textup{GHZ}^n}\equiv 
\left(\ket{0}^{\otimes n}+\ket{1}^{\otimes n}\right)/\sqrt{2}$ denotes the maximally-entangled Greenberger-Horne-Zeilinger (GHZ) states on $n$ qubits, and $\neigh_1 = \{1,2,3\},$ $\neigh_2=\{2,3,4\}$. 
This shows that we can stabilize states that are {\em arbitrarily close} to states that are provably not QLS, 
as the GHZ states \cite{ticozzi-ql1}, thereby achieving practical stabilization of the latter.
\end{enumerate}

\section{Conclusions}

We have introduced alternating projection methods based on sequences of CPTP projections, and used 
them in designing discrete-time stabilizing dynamics for entangled states in multipartite quantum systems subject 
to realistic quasi-locality constraints. 
When feasible, pursuing stabilization instead of preparation offers important 
advantages, including the possibility to retrieve the target state {\em on-demand}, at any (discrete) time after 
sufficient convergence is attained, since the invariance of the latter ensures that it is not ruined by subsequent maps.
We show that the proposed methods are also suitable for distributed, randomized and unsupervised implementations 
on large networks. While the locality constraints we impose on the discrete-time dynamics are stricter, the stabilizable 
states are, remarkably, the same that are stabilizable for continuous-time frustration-free generators. 
  
From a methodological standpoint, our results shed further light on the structure and {\em intersection of fixed-point sets} of CPTP maps, a structure of interest not only in control, but also in operator-algebraic approaches to quantum systems \cite{bratteli}, quantum statistics \cite{petz-book} and quantum error correction theory \cite{viola-generalnoise,viola-IPS, ticozzi-isometries}.  In particular, we show that the intersection of fixed-point sets is {\em still} a fixed-point set, as long as it contains a full-rank state. In developing our results, we use both standard results from classical alternating projections and Lyapunov methods tailored to the positive linear maps at hand. 

Towards applications, the proposed alternating projection methods are in principle suitable for implementation in digital 
open-quantum system simulators, such as demonstrated in proof-of-principle trapped-ion experiments \cite{barreiro}. 
Beside providing protocols for stabilizing relevant classes of pure entangles states, our methods point to 
an alternative approach for constructing quantum samplers using quasi-local resources.

Some developments of this line of research are worth highlighting. First, in order to extend the applicability of the proposed methods to more general classes of states, as well as to establish a tighter link to quantum error correction and 
dissipative code preparation, it is natural to look at discrete-time {\em conditional stabilization}, in the spirit of 
\cite{ticozzi-ql1}. Notably, in \cite{ticozzi-newqconsensus}, it has been shown that GHZ states and {\em all} 
Dicke states can be made conditionally 
asymptotically stable for  QL discrete-time dynamics, with a suitable basin of attraction.
Second, while we recalled some basic classical bounds on the convergence speed,  
that apply to the stabilization of full-rank states, their geometric nature makes it hard to obtain useful insight from them.
A more intuitive approach to convergence speed and its optimization, following e.g. \cite{ticozzi-NV,cirillo-decompositions}, may offer a more promising venue in that respect. It has also been recently shown that linear Lyapunov functions can not only be used to prove convergence, but also provide sharp bounds on the convergence speed in continuous-time dynamics \cite{tristan}. It would be interesting to extend the analysis to the non-homogeneous, discrete-time cases considered in this work.  Lastly, the characterization of physically relevant scenarios in which {\em finite-time stabilization} is possible under 
locality constraints is a challenging open problem, which we plan to address elsewhere \cite{fts-paper}.


\section*{Acknowledgements}

It is a pleasure to acknowledge stimulating discussions on the topics of this work with A. Ferrante e L. Finesso. F.T. 
is especially grateful to V. Umanit\`a and E. Sasso for pointing him towards Takesaki's theorem.  
Work at Dartmouth was supported by the National Science Foundation through grant No. PHY-1620541.

\appendix

\subsection{Angles between subspaces}
\label{app:cosine}

Define the function $\arccos:[-1,1]\rightarrow [-\frac{\pi}{2},\frac{\pi}{2} ]$. We will use only the elements in interval $[0,1]$. Then the {\em angle} $\theta(\mathcal{M},\mathcal{N})$ between two closed subspaces $\mathcal{M}$ and 
$\mathcal{N}$ of $\hilbert$ is 
an element of $[0,\frac{\pi}{2}]$. We have the following:
{\defin The cosine $c(\mathcal{M},\mathcal{N})$ between two closed subspaces $\mathcal{M}$ and $\mathcal{N}$ of 
$\hilbert$ is given by 
\beqan 
c(\mathcal{M},\mathcal{N})&\equiv &\sup \Big\{|\langle x,y\rangle|:x\in \mathcal{M}\cap (\mathcal{M}\cap \mathcal{N})^\perp,\\ &&\|x\|\leq1, y\in \mathcal{N}\cap(\mathcal{M}\cap \mathcal{N})^\perp, \|y\|\leq1 \Big\}.
\eeqan 
Then the angle is given by: 
$$\theta(\mathcal{M},\mathcal{N})=\arccos(c(\mathcal{M},\mathcal{N})).$$}
Some consequences of the above definitions are the following:
\begin{enumerate}
\item $0\leq c(\mathcal{M},\mathcal{N})\leq 1$;
\item $c(\mathcal{M},\mathcal{N})=c(\mathcal{N},\mathcal{M})$;
\item  $c(\mathcal{M},\mathcal{N})=\|P_\mathcal{M}P_\mathcal{N}-P_{\mathcal{M}\cap \mathcal{N}}\|=\|P_\mathcal{M}P_\mathcal{N}P_{(\mathcal{M}\cap \mathcal{N})^\perp}\|$.
\end{enumerate} 
\noindent 
We next state the result that gives the exact rate in case of projection onto two subspaces \cite{escalante}:
{\thm In the norm induced by the inner product, and for each $n$, the following equality holds:
$$\|(P_{\mathcal{M}_2}P_{\mathcal{M}_1})^n-P_{\mathcal{M}_1\cap \mathcal{M}_2}\|= c(\mathcal{M}_1,\mathcal{M}_2)^{2n-1}.$$ }
In case of alternating projections on the intersection of more than two subspaces, an exact expression is no longer 
available, however an upper bound may be given \cite{escalante}:
{\thm \label{thm:conv_n} For each $i=1,2,\ldots,r$, let $\mathcal{M}_i$ be a closed subspace of $\hilbert$. 
Then, for each $x\in \hilbert$, and for any integer $n\geq1$ it holds:
$$\|(P_{\mathcal{M}_r}...P_{\mathcal{M}_1})^n x-P_{\bigcap_{i=1}^r\mathcal{M}_i}x\|\leq c^{\frac{n}{2}}\| 
x-P_{\bigcap_{i=1}^r\mathcal{M}_i}x\|,$$
where the contraction coefficient
$$c=1-\prod_{i=1}^{r-1}\sin^2\theta_i,$$
and $\theta_i$ is the angle between $\mathcal{M}_i$ and $\bigcap_{j=i+1}^r\mathcal{M}_j$. }

\subsection{Non-orthogonality of ${\cal E}_{\cal A}$ with respect to the Hilbert-Schmidt inner product}
\label{app1}

Let us decompose a full-rank fixed point set $\mathcal{A}_\rho=\bigoplus_\ell \mathcal{A}_\ell=
\bigoplus_\ell \mathcal{B}(\hilbert_{S,\ell})\otimes \tau_\ell$, (where $\tau_\ell\equiv  \tau_{F,\ell} $).  
By definition, the {\em orthogonal} projection of $X$ onto $\mathcal{A}_i$ is given by 
$$P_\mathcal{A}(X) \equiv \sum_{\ell,i} 
\langle\sigma_{\ell,i }\otimes { \tau_\ell} ,X \rangle_{HS} \,
\sigma_{\ell,i}\otimes \tau_\ell, $$
where $\sigma_{\ell,i}\otimes \tau_\ell$ is an orthonormal basis for $\mathcal{A}_\ell$. Note that the outcome only depends on the restrictions of $X$ to the supports of the $\Ai_\ell.$ Hence, decompose $X \equiv \sum_\ell X_\ell+\Delta X$, where $X_\ell=\Pi_{SF,\ell}X\Pi_{SF,\ell},$ 
and  
further decompose $X_\ell \equiv \sum_k A_{\ell,k}\otimes B_{\ell,k},$ so we can write:
\begin{eqnarray*}
 P_\mathcal{A}(X)&\hspace*{-2mm}=\hspace*{-2mm}&\bigoplus_i\sum_{j,\ell}\Big(\sum_k \trace[(\sigma_j\otimes \tau_\ell)(A_{\ell,k}\otimes B_{\ell,k})]\sigma_j\otimes\tau_\ell \Big)\\
&\hspace*{-2mm}=\hspace*{-2mm}&\bigoplus_\ell\sum_{j,\ell}\Big(\trace[\sigma_j \sum_k(A_{\ell,k} \trace(\tau_\ell B_{\ell,k}))]\sigma_j\otimes\tau_\ell)\Big).
\end{eqnarray*}
By comparing the latter equation with Eq. \eqref{eq:proj_2}, we have that $P_\mathcal{A}=\Ei_\Ai$ if and only if  $\sum_k(A_k \trace(\tau_jB_k))=\trace_{F,\ell}(X_\ell),$ which is equivalent to request that $\tau_j=\lambda_\ell I.$ Hence, unless $\cal{A}_\rho$ contains the completely mixed state, ${\mathcal{E}}_\Ai$ in Eq. \eqref{eq:proj_2} is not an orthogonal projection with respect to the Hilbert-Schmidt inner product. \qed

\bibliographystyle{IEEEtran}
\bibliography{bibQL}

\end{document}